\documentclass[aps,prl,showpacs,preprintnumbers,%
               twocolumn,groupedaddress,amsfonts]{revtex4}
\usepackage{graphicx}
\usepackage{dcolumn}
\usepackage{bm}
\suppressfloats[!]
\begin{document}
\preprint{Fermilab-Pub-05/523-E}
\title{\boldmath Measurement of the Isolated Photon Cross Section
in $p\bar{p}$ Collisions at $\sqrt{s}=1.96$~TeV}
%
\author{                                                                      
V.M.~Abazov,$^{36}$                                                           
B.~Abbott,$^{75}$                                                             
M.~Abolins,$^{65}$                                                            
B.S.~Acharya,$^{29}$                                                          
M.~Adams,$^{52}$                                                              
T.~Adams,$^{50}$                                                              
M.~Agelou,$^{18}$                                                             
J.-L.~Agram,$^{19}$                                                           
S.H.~Ahn,$^{31}$                                                              
M.~Ahsan,$^{59}$
% ...
\\                                                                            
\vskip 0.50cm                                                                 
\centerline{(D\O\ Collaboration)}                                             
\vskip 0.50cm                                                                 
}                                                                             
\affiliation{                                                                 
\centerline{$^{1}$Universidad de Buenos Aires, Buenos Aires, Argentina}       
\centerline{$^{2}$LAFEX, Centro Brasileiro de Pesquisas F{\'\i}sicas,         
                  Rio de Janeiro, Brazil}                                     
\centerline{$^{3}$Universidade do Estado do Rio de Janeiro,                   
                  Rio de Janeiro, Brazil}                                     
\centerline{$^{4}$Instituto de F\'{\i}sica Te\'orica, Universidade            
                  Estadual Paulista, S\~ao Paulo, Brazil}                     
\centerline{$^{5}$University of Alberta, Edmonton, Alberta, Canada,           
               Simon Fraser University, Burnaby, British Columbia, Canada,}   
\centerline{York University, Toronto, Ontario, Canada, and                    
         McGill University, Montreal, Quebec, Canada}                         
\centerline{$^{6}$Institute of High Energy Physics, Beijing,                  
                  People's Republic of China}                                 
\centerline{$^{7}$University of Science and Technology of China, Hefei,       
                  People's Republic of China}                                 
\centerline{$^{8}$Universidad de los Andes, Bogot\'{a}, Colombia}             
\centerline{$^{9}$Center for Particle Physics, Charles University,            
                  Prague, Czech Republic}                                     
\centerline{$^{10}$Czech Technical University, Prague, Czech Republic}        
\centerline{$^{11}$Center for Particle Physics, Institute of Physics,         
                   Academy of Sciences of the Czech Republic,                 
                   Prague, Czech Republic}                                    
\centerline{$^{12}$Universidad San Francisco de Quito, Quito, Ecuador}        
\centerline{$^{13}$Laboratoire de Physique Corpusculaire, IN2P3-CNRS,         
                  Universit\'e Blaise Pascal, Clermont-Ferrand, France}       
\centerline{$^{14}$Laboratoire de Physique Subatomique et de Cosmologie,      
                  IN2P3-CNRS, Universite de Grenoble 1, Grenoble, France}     
\centerline{$^{15}$CPPM, IN2P3-CNRS, Universit\'e de la M\'editerran\'ee,     
                  Marseille, France}                                          
\centerline{$^{16}$IN2P3-CNRS, Laboratoire de l'Acc\'el\'erateur              
                  Lin\'eaire, Orsay, France}                                  
\centerline{$^{17}$LPNHE, IN2P3-CNRS, Universit\'es Paris VI and VII,         
                  Paris, France}                                              
\centerline{$^{18}$DAPNIA/Service de Physique des Particules, CEA, Saclay,    
                  France}                                                     
\centerline{$^{19}$IReS, IN2P3-CNRS, Universit\'e Louis Pasteur, Strasbourg,  
                France, and Universit\'e de Haute Alsace, Mulhouse, France}   
\centerline{$^{20}$Institut de Physique Nucl\'eaire de Lyon, IN2P3-CNRS,      
                   Universit\'e Claude Bernard, Villeurbanne, France}         
\centerline{$^{21}$III. Physikalisches Institut A, RWTH Aachen,               
                   Aachen, Germany}                                           
\centerline{$^{22}$Physikalisches Institut, Universit{\"a}t Bonn,             
                  Bonn, Germany}                                              
\centerline{$^{23}$Physikalisches Institut, Universit{\"a}t Freiburg,         
                  Freiburg, Germany}                                          
\centerline{$^{24}$Institut f{\"u}r Physik, Universit{\"a}t Mainz,            
                  Mainz, Germany}                                             
\centerline{$^{25}$Ludwig-Maximilians-Universit{\"a}t M{\"u}nchen,            
                   M{\"u}nchen, Germany}                                      
\centerline{$^{26}$Fachbereich Physik, University of Wuppertal,               
                   Wuppertal, Germany}                                        
\centerline{$^{27}$Panjab University, Chandigarh, India}                      
\centerline{$^{28}$Delhi University, Delhi, India}                            
\centerline{$^{29}$Tata Institute of Fundamental Research, Mumbai, India}     
\centerline{$^{30}$University College Dublin, Dublin, Ireland}                
\centerline{$^{31}$Korea Detector Laboratory, Korea University,               
                   Seoul, Korea}                                              
\centerline{$^{32}$SungKyunKwan University, Suwon, Korea}                     
\centerline{$^{33}$CINVESTAV, Mexico City, Mexico}                            
\centerline{$^{34}$FOM-Institute NIKHEF and University of                     
                  Amsterdam/NIKHEF, Amsterdam, The Netherlands}               
\centerline{$^{35}$Radboud University Nijmegen/NIKHEF, Nijmegen, The          
                  Netherlands}                                                
\centerline{$^{36}$Joint Institute for Nuclear Research, Dubna, Russia}       
\centerline{$^{37}$Institute for Theoretical and Experimental Physics,        
                  Moscow, Russia}                                             
\centerline{$^{38}$Moscow State University, Moscow, Russia}                   
\centerline{$^{39}$Institute for High Energy Physics, Protvino, Russia}       
\centerline{$^{40}$Petersburg Nuclear Physics Institute,                      
                   St. Petersburg, Russia}                                    
\centerline{$^{41}$Lund University, Lund, Sweden, Royal Institute of          
                   Technology and Stockholm University, Stockholm,            
                   Sweden, and}                                               
\centerline{Uppsala University, Uppsala, Sweden}                              
\centerline{$^{42}$Physik Institut der Universit{\"a}t Z{\"u}rich,            
                    Z{\"u}rich, Switzerland}                                  
\centerline{$^{43}$Lancaster University, Lancaster, United Kingdom}           
\centerline{$^{44}$Imperial College, London, United Kingdom}                  
\centerline{$^{45}$University of Manchester, Manchester, United Kingdom}      
\centerline{$^{46}$University of Arizona, Tucson, Arizona 85721, USA}         
\centerline{$^{47}$Lawrence Berkeley National Laboratory and University of    
                  California, Berkeley, California 94720, USA}                
\centerline{$^{48}$California State University, Fresno, California 93740, USA}
\centerline{$^{49}$University of California, Riverside, California 92521, USA}
\centerline{$^{50}$Florida State University, Tallahassee, Florida 32306, USA} 
\centerline{$^{51}$Fermi National Accelerator Laboratory, Batavia,            
                   Illinois 60510, USA}                                       
\centerline{$^{52}$University of Illinois at Chicago, Chicago,                
                   Illinois 60607, USA}                                       
\centerline{$^{53}$Northern Illinois University, DeKalb, Illinois 60115, USA} 
\centerline{$^{54}$Northwestern University, Evanston, Illinois 60208, USA}    
\centerline{$^{55}$Indiana University, Bloomington, Indiana 47405, USA}       
\centerline{$^{56}$University of Notre Dame, Notre Dame, Indiana 46556, USA}  
\centerline{$^{57}$Iowa State University, Ames, Iowa 50011, USA}              
\centerline{$^{58}$University of Kansas, Lawrence, Kansas 66045, USA}         
\centerline{$^{59}$Kansas State University, Manhattan, Kansas 66506, USA}     
\centerline{$^{60}$Louisiana Tech University, Ruston, Louisiana 71272, USA}   
\centerline{$^{61}$University of Maryland, College Park, Maryland 20742, USA} 
\centerline{$^{62}$Boston University, Boston, Massachusetts 02215, USA}       
\centerline{$^{63}$Northeastern University, Boston, Massachusetts 02115, USA} 
\centerline{$^{64}$University of Michigan, Ann Arbor, Michigan 48109, USA}    
\centerline{$^{65}$Michigan State University, East Lansing, Michigan 48824,   
                   USA}                                                       
\centerline{$^{66}$University of Mississippi, University, Mississippi 38677,  
                   USA}                                                       
\centerline{$^{67}$University of Nebraska, Lincoln, Nebraska 68588, USA}      
\centerline{$^{68}$Princeton University, Princeton, New Jersey 08544, USA}    
\centerline{$^{69}$State University of New York, Buffalo, New York 14260, USA}
\centerline{$^{70}$Columbia University, New York, New York 10027, USA}        
\centerline{$^{71}$University of Rochester, Rochester, New York 14627, USA}   
\centerline{$^{72}$State University of New York, Stony Brook,                 
                   New York 11794, USA}                                       
\centerline{$^{73}$Brookhaven National Laboratory, Upton, New York 11973, USA}
\centerline{$^{74}$Langston University, Langston, Oklahoma 73050, USA}        
\centerline{$^{75}$University of Oklahoma, Norman, Oklahoma 73019, USA}       
\centerline{$^{76}$Oklahoma State University, Stillwater, Oklahoma 74078, USA}
\centerline{$^{77}$Brown University, Providence, Rhode Island 02912, USA}     
\centerline{$^{78}$University of Texas, Arlington, Texas 76019, USA}          
\centerline{$^{79}$Southern Methodist University, Dallas, Texas 75275, USA}   
\centerline{$^{80}$Rice University, Houston, Texas 77005, USA}                
\centerline{$^{81}$University of Virginia, Charlottesville, Virginia 22901,   
                   USA}                                                       
\centerline{$^{82}$University of Washington, Seattle, Washington 98195, USA}  
}                                                                             

\begin{abstract}\noindent
The cross section for the inclusive production of isolated photons has
been measured in $p\bar{p}$ collisions at $\sqrt{s} = 1.96$ TeV with
the D\O\ detector at the Fermilab Tevatron Collider. The photons span
transverse momenta $23$ to $300$~GeV and have pseudorapidity
$|\eta|<0.9$.  The cross section is compared with the results from two
next-to-leading order perturbative QCD calculations.  The theoretical
predictions agree with the measurement within uncertainties.
\end{abstract}
\pacs{13.85.Qk, 12.38.Qk}
\date{\today}
\maketitle
Photons originating in the hard interaction between two partons are
typically produced in hadron collisions via quark-gluon Compton
scattering or quark--anti-quark
annihilation~\cite{owens,aurenche,berger,Baur:2000xd}.  Studies of
these direct photons with large transverse momenta, $p_T^\gamma$,
provide precision tests of perturbative QCD (pQCD) as well as
information on the distribution of partons within protons,
particularly the gluon.  These data were used in global fits of parton
distributions functions (PDFs) and complement analyses of deep
inelastic scattering, Drell-Yan pair production, and jet
production~\cite{Fits}.  Photons from energetic $\pi^0$ and $\eta$
mesons are the main background to direct photon production especially
at small $p_T^\gamma$~\cite{Apanasevich:2000eq}.  Since these mesons
are produced inside jets, their contribution can be suppressed with
respect to direct photons by requiring the photon be isolated from
other particles.  Isolated electrons from the electroweak production
of $W$ and $Z$ bosons also contribute to the background at high
$p_T^\gamma$.  Previous measurements of photon production at hadron
colliders successfully used these isolation techniques to extract the
photon
signal~\cite{Albajar:1988im,Alitti:1992hn,Abe:1994rr,Abbott:1999kd,Abazov:2001af,Acosta:2002ya,Acosta:2004bg}.

We present, in this Letter, a measurement of the cross section for the
inclusive production of isolated photons with pseudorapidity
$|\eta|<0.9$ in $p\bar{p}$ collisions at $\sqrt{s}=1.96$~TeV.
(Pseudorapidity is defined as $\eta=-\ln\tan(\theta/2)$, where
$\theta$ is the polar angle with respect to the proton beam
direction.)  The data sample corresponds to an integrated luminosity
${ L}=326 \pm 21$~pb$^{-1}$~\cite{luminosity} accumulated in
2002--2004 with the D\O\ detector~\cite{D0_det_run2} at the Fermilab
Tevatron Collider.  The primary tool for photon detection is the
central part of a liquid-argon and uranium calorimeter covering
$|\eta|<1.1$. Two additional calorimeters, housed in separate
cryostats, extend the coverage to $|\eta|<4.2$~\cite{D0_det_run1}.
The electromagnetic section of the central calorimeter (EM) is
segmented longitudinally into four layers (EM1$-$EM4) of 2, 2, 7, and
10 radiation lengths, respectively, and transversely into cells in
$\eta$ and azimuthal angle, $\Delta\eta\times\Delta\phi=0.1\times0.1$
($0.05\times0.05$ in the EM3 layer at the electromagnetic shower
maximum), yielding a good angular resolution for photons and
electrons.  The calorimeter surrounds a preshower detector and a
tracking system which consists of silicon microstrip and scintillating
fiber trackers ($0.3$ radiation lengths) located within a 2~T
solenoidal magnet.  The total amount of material between the
interaction point and the first active layer of the calorimeter is
equivalent to approximately $3.5-4.5$~radiation lengths (increasing
with $|\eta|$).  The position and width of the $Z$~boson mass peak
were used to determine the EM calorimeter calibration factors and the
EM energy resolution~\cite{Zgamma}.

Photon candidates were formed from clusters of calorimeter cells
within a cone of radius ${\mathcal R}=\sqrt{(\Delta\eta)^2+(\Delta\phi)^2}=0.4$; the energy was then
recalculated from the inner core with ${\mathcal R}=0.2$.  Candidates
were selected if there was significant energy in the EM calorimeter
layers ($>95\%$), and the probability to have a spatially-matched track was less
than $0.1\%$, and they satisfied the isolation requirement
$(E_{total}(0.4)-E_{EM}(0.2))/E_{EM}(0.2)<0.10$, where
$E_{total}(0.4)$ is the total energy in a cone with ${\mathcal R}=0.4$ and
$E_{EM}(0.2)$ is the EM energy within ${\mathcal R}=0.2$.  Photon candidates with
energy measurements biased by calorimeter module boundaries and
structures were removed from consideration; the geometric acceptance
was $A=(84.2\pm1.5)\%$.  Potential backgrounds from cosmic rays and
leptonic $W$ boson decays were suppressed by requiring the missing
transverse energy, calculated from the vector sum of the transverse
energies of calorimeter cells, to be less than $0.7p_T^\gamma$.  The
efficiency for the above requirements was estimated with direct
photons generated by {\sc pythia}~\cite{pythia}. Events were processed
with the {\sc geant} detector simulation package and overlaid with
detector noise and minimum bias interactions~\cite{D0_det_run2}.  The
efficiency (excluding acceptance) rose from $(82\pm5)\%$ at
$p_T^\gamma\approx24$~GeV to a plateau of $(92\pm3)\%$ at $p_T^\gamma>110$~GeV.  
We used $Z\rightarrow e^+e^-$ events~\cite{Zgamma}, due
to the similarity between electron- and photon-initiated showers, 
to verify the selection efficiencies estimated with the Monte
Carlo simulation (MC).  The photon sample was acquired with a
three-level trigger system that relied on hardware signals from the
calorimeter and fast, software-based, photon reconstruction.  The
trigger was $(71\pm9)\%$ efficient for photon candidates with
$p_T^\gamma\approx24$~GeV, $(93\pm2)\%$ at $p_T^\gamma\approx32$~GeV
and greater than $98\%$ for $p_T^\gamma>40$~GeV.  Every event was
required to have a vertex, reconstructed with at least three tracks,
within 50~cm of the nominal center of the detector along the beam
axis; the efficiency for this requirement ranged from $(90.0\pm0.3)\%$
to $(95.3\pm0.1)\%$ as a function of instantaneous luminosity.

\begin{figure}
\centering
\includegraphics[scale=0.44,bb=10 30 500 500,clip=true]{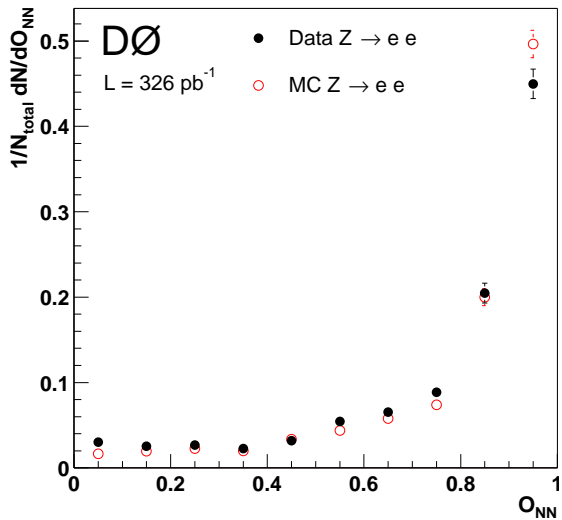} 
\caption{Normalized distributions of NN
output ($O_{\rm NN}$) in $Z\rightarrow e^+e^-$ events for
data~($\bullet$) and MC~($\circ$).
\label{fig:ANN_GamEmjright}}
\end{figure}

Four variables were used to further suppress the background: the
number of EM1 cells with energy greater than $400$~MeV within
${\mathcal R}<0.2$ and within $0.2<{\mathcal R}<0.4$, the scalar sum
of the transverse momenta of tracks within $0.05<{\mathcal R}<0.4$,
and the energy-weighted cluster width in the finely-segmented EM3
layer.  These variables were input to an artificial neural network
(NN), built with the {\sc jetnet} package~\cite{jetnet}, to suppress
background and to estimate the purity of the resulting photon sample.
The NN was trained to discriminate between direct photons and
background events. The background events, produced with QCD and
electroweak processes in {\sc pythia}, were preselected with loose
criteria to increase statistics and to exclude high-momentum
bremsstrahlung photons produced from partons.  The resulting NN
output, $O_{\rm NN}$, peaks at unity for signal events and at zero for
background events.  Events with $O_{\rm NN}>0.5$ were considered in
this analysis, yielding a high photon selection efficiency of
$(93.7\pm0.2)\%$ and good background rejection.  
The NN was tested in MC and data using electrons from $Z$ boson decays; 
the resulting $O_{\rm NN}$ distributions are shown in
Fig.~\ref{fig:ANN_GamEmjright}.  The systematic uncertainty on the
signal efficiency for the $O_{\rm NN}$ requirement, estimated with
electrons 
from the $Z$ boson samples, is $2.4\%$.

\begin{figure}
\centering
\includegraphics[scale=0.44, bb=10 30 500 500,clip=true]{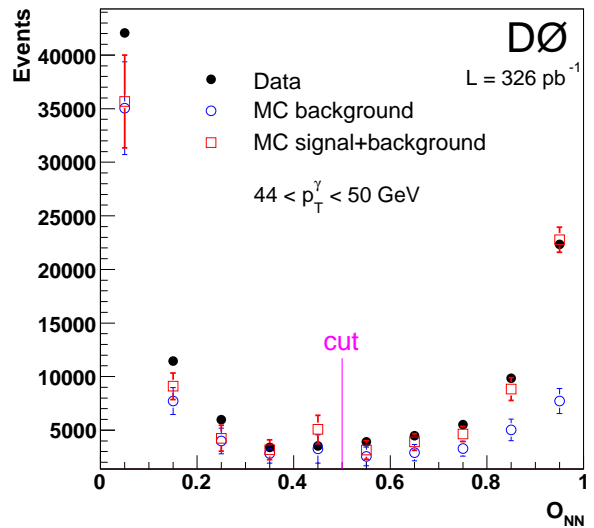}
\caption{
Distribution of the number of events in data~($\bullet$) as a function
of the NN output ($O_{\rm NN}$) for $44<p_T^\gamma<50$ GeV.  The
contributions from MC background~($\circ$) and summed MC signal and
background~($_{^\Box}$) are also shown.  The MC points were weighted
according to the fitted purity (the errors shown are statistical).
\label{fig:pur_exam1}}
\end{figure}

\begin{figure}
\centering
\includegraphics[scale=0.44, bb=30 30 515 495,clip=true]{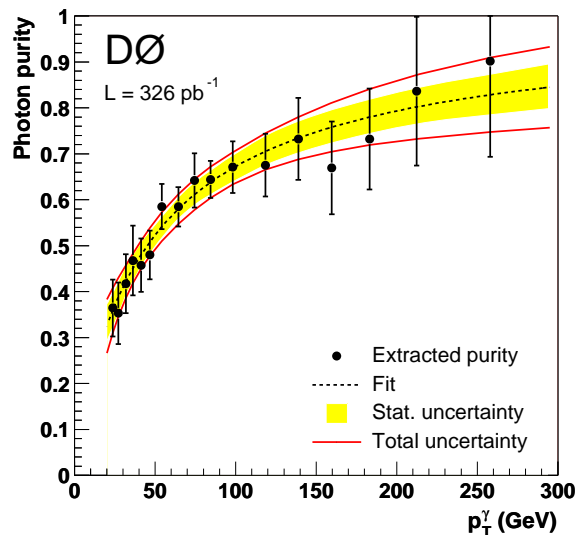}
\caption{Dependence of the photon purity on $p_T^\gamma$. 
The dashed line represents a fit to these points, the filled area
corresponds to the statistical uncertainty band, and the solid lines
to the total uncertainty band.
The NN output in data was fit to the shapes of the
MC signal and background samples.
\label{fig:pur_tot}}
\end{figure}

The photon purity ($\mathcal P$), defined as the ratio of signal to
signal plus background, was determined statistically for each
$p_T^\gamma$ bin.  Distributions of the number of events as a function
of $O_{\rm NN}$ are shown for data and MC in Fig.~\ref{fig:pur_exam1}
for the $44<p_T^\gamma<50$ GeV interval.  The MC events in this figure
were weighted by the fractions that resulted from 
the fit performed with the {\sc hmcmll} package~\cite{HMCMLL}.  
The data are well described by the sum of MC signal and
background samples, especially for events with $O_{\rm NN}>0.5$.
Photon purities are shown in Fig.~\ref{fig:pur_tot} as a function of
$p_T^\gamma$.  The purity uncertainty is dominated by MC statistics at
low $p_T^\gamma$ and data statistics at high $p_T^\gamma$.  Systematic
uncertainties were estimated by using two alternate fitting functions
and by varying the number of bins used in the {\sc hmcmll} fits.
The {\sc pythia} fragmentation model was an additional source of
systematic uncertainty.  This uncertainty was estimated by varying the
production rate of $\pi^0$, $\eta$, $K^0_s$, and $\omega$ mesons by
$\pm50\%$~\cite{Monique_pion_Sjostr} resulting in an uncertainty of
$7.5\%$ at $p_T^\gamma\approx24$~GeV, $2\%$ at
$p_T^\gamma\approx50$~GeV, and $1\%$ for $p_T^\gamma>70$~GeV.

\begin{figure}
\centering
\includegraphics[scale=0.44,bb=5 25 515 495,clip=true]{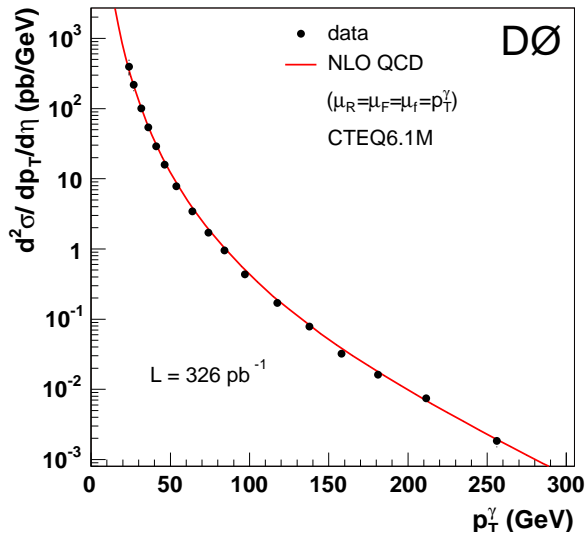}
\caption{
The inclusive cross section for the production of isolated photons as
a function of $p_T^\gamma$.  The results from the NLO pQCD calculation
with {\sc jetphox} are shown as solid line.
\label{fig:DTleft}}
\end{figure}

The isolated-photon cross section is measured using the following
definition:
\begin{eqnarray}
\frac{d^2\sigma}{dp_Td\eta} = \frac{N\;{\mathcal P}\;U}
{{ L}\;\Delta p_T^\gamma\Delta\eta\;A\epsilon}
\label{eq:cross}
\end{eqnarray}
where $N$ is the number of photon candidates, $\epsilon$ is the
combined efficiency for the selection criteria described above, and
$\Delta p_T^\gamma$ and $\Delta\eta$ are the bin sizes. The factor $U$
corrects the cross section for the effects of the finite resolution of
the calorimeter.  This unsmearing was performed, as a function of
$p_T^\gamma$, by iteratively fitting the convolution of an ansatz
function with an energy resolution function.  The uncertainty in this
correction was estimated using two different ansatz functions and
included the uncertainty in the energy resolution.  An additional
correction was applied to $p_T^\gamma$ for the difference in the
energy deposited in the material upstream of the calorimeter between
electrons (used for the energy calibration) and photons.  This
correction to $p_T^\gamma$ was approximately $1.9\%$ at $20$~GeV,
$1.0\%$ at $40$~GeV, and less than $0.3\%$ for $p_T^\gamma>70$~GeV.
The measured cross section, together with statistical and systematic
uncertainties, is presented in Fig.~\ref{fig:DTleft} and
Table~\ref{tab:cross}.  (The data points are plotted at the $p_T$
value for which a smooth function describing the cross section is
equal to the average cross section in the bin~\cite{TW}.)  Sources of
systematic uncertainty include luminosity ($6.5\%$), event vertex
determination ($3.6\%-5.0\%$), energy calibration ($9.6\%-5.5\%$), the
fragmentation model ($7.3\%-1.0\%$), photon conversions ($3\%$), and
the photon purity fit uncertainty (shown in Fig.~\ref{fig:pur_tot}) as
well as statistical uncertainties on the determination of geometrical
acceptance ($1.5\%$), trigger efficiency ($11\%-1\%$), selection
efficiency ($5.4\%-3.8\%$) and unsmearing ($1.5\%$).  The uncertainty
ranges above are quoted with the uncertainty at low $p_T^\gamma$ first
and the uncertainty at high $p_T^\gamma$ second.  Most of these
systematic uncertainties have large ($>80\%$) bin-to-bin correlations
in $p_T^\gamma$.  Varying the choice of NN cut from 0.3 to~0.7 changed
the measured cross section by less than $5\%$.
\begin{figure}
\centering
\includegraphics[scale=0.44,bb=20 25 515 495,clip=true]{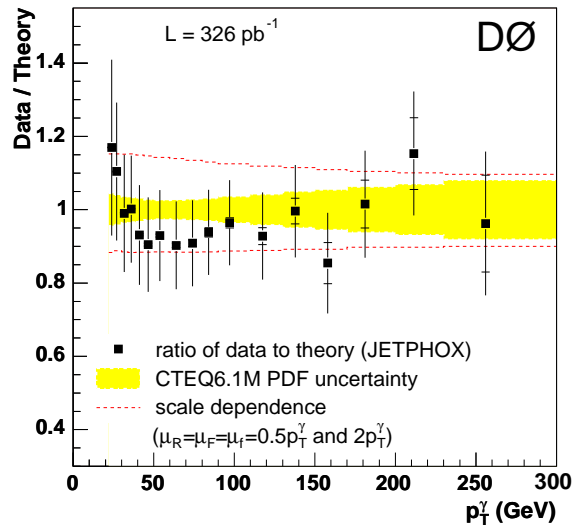}
\caption{The ratio of the measured
cross section to the theoretical predictions from {\sc jetphox}. The
full vertical lines correspond to the overall uncertainty while the
internal line indicates just the statistical uncertainty. Dashed lines
represents the change in the cross section when varying the
theoretical scales by factors of two.  The shaded region indicates the
uncertainty in the cross section estimated with CTEQ6.1 PDFs.
\label{fig:DTright}}
\end{figure}
\begin{table}
\centering
\caption{
The measured differential cross section for the production of isolated
photons, averaged over $|\eta|<0.9$, in bins of $p_T^\gamma$. $\langle
p_T^\gamma\rangle$ is the average $p_T^\gamma$ within each bin. The
columns $\delta\sigma_{\rm stat}$ and $\delta\sigma_{\rm syst}$
represent the statistical and systematic uncertainties respectively.
(Five events with $p_T^\gamma>300$ GeV, including one with
$p_T^\gamma=442$~GeV, were not considered in this analysis.)
\label{tab:cross}}
\begin{tabular}{r@{$-$}lcr@{$\times$}lccc} \hline\hline
\multicolumn{2}{c}{$p_T^\gamma$} 
& $\;\;$ $\langle p_T^\gamma\rangle$ $\;\;$
& \multicolumn{2}{c}{$d^2\sigma/dp_T^\gamma d\eta$ $\;\;$ } 
& $\;\;$ $\delta\sigma_{\rm stat}$ $\;\;$
& $\delta\sigma_{\rm syst}$ $\;\;$ \\
\multicolumn{2}{c}{(GeV)}   
& $\;\;$ (GeV)  $\;\;\;\;$ 
&\multicolumn{2}{c}{$\!\!\!\!$(pb/GeV) } 
& $\;\;$ (\%) $\;\;$
&  (\%) \\
\hline
     23 &  25 &  23.9 & 4.14 & $ 10^2$ &    0.1 & 23  \\ 
     25 &  30 &  26.9 & 2.21 & $ 10^2$ &    0.1 & 19  \\ 
     30 &  34 &  31.7 & 1.01 & $ 10^2$ &    0.2 & 16  \\ 
     34 &  39 &  36.0 & 5.37 & $ 10^1$ &    0.2 & 15  \\ 
     39 &  44 &  41.1 & 2.88 & $ 10^1$ &    0.3 & 14  \\ 
     44 &  50 &  46.5 & 1.58 & $ 10^1$ &    0.4 & 13  \\ 
     50 &  60 &  53.8 & 7.90 & $ 10^0$ &    0.4 & 13  \\ 
     60 &  70 &  63.9 & 3.39 & $ 10^0$ &    0.6 & 13  \\
     70 &  80 &  74.1 & 1.68 & $ 10^0$ &    0.9 & 12  \\ 
     80 &  90 &  84.1 & 9.34 & $ 10^{-1}$ & 1.3 & 12  \\ 
     90 & 110 &  97.2 & 4.38 & $ 10^{-1}$ & 1.4 & 12  \\ 
    110 & 130 & 118   & 1.66 & $ 10^{-1}$ & 2.3 & 12  \\
    130 & 150 & 138   & 7.61 & $ 10^{-2}$ & 3.5 & 13  \\
    150 & 170 & 158   & 3.20 & $ 10^{-2}$ & 5.6 & 13  \\ 
    170 & 200 & 181   & 1.59 & $ 10^{-2}$ & 6.5 & 14  \\
    200 & 230 & 212   & 7.36 & $ 10^{-3}$ & 9.8 & 14  \\  
    230 & 300 & 256   & 1.81 & $ 10^{-3}$ & 13  & 15  \\
\hline\hline
\end{tabular}
\end{table}

Results from a next-to-leading order (NLO) pQCD calculation ({\sc
jetphox}~\cite{DIPHOX,JETPHOX}) are compared to our measured cross
section in Fig.~\ref{fig:DTleft}.  These results were derived using
the CTEQ6.1M~\cite{cteq61} PDFs and the BFG~\cite{BFG} fragmentation
functions (FFs).  The renormalization, factorization, and
fragmentation scales were chosen to be
$\mu_{R}\!=\!\mu_{F}\!=\!\mu_{f}\!=\!p_T^\gamma$. Another NLO pQCD
calculation~\cite{GV}, based on the small-cone approximation and
utilizing different FFs~\cite{GRV}, gave consistent results (within
$4\%$).  As shown in Fig.~\ref{fig:DTright}, the calculation agrees,
within uncertainties, with the measured cross section.  The scale
dependence in the NLO pQCD theory, estimated by varying scales by
factors of two, are displayed in Fig.~\ref{fig:DTright} as dashed
lines.  The span of these results is comparable to the overall
uncertainty in the cross section measurement.  The filled area in
Fig.~\ref{fig:DTright} represents the uncertainty associated with the
CTEQ6.1M PDFs.  The central values of the predictions change by less
than 7\% when the PDFs are replaced by MRST2004~\cite{Martin:2004ir}
or Alekhin2004~\cite{Alekhin:2002fv}.  The calculation is also
sensitive to the implementation of the isolation requirements
including the hadronic fraction in the ${\mathcal R}=0.2$ cone around
the photon.  The variation in the predicted cross section for $50\%$
changes in the cut values for these criteria was found to be less than
$3\%$~\cite{VV_comm}.

In conclusion, we have measured the cross section for the production
of isolated photons with $|\eta|<0.9$ produced in $p\bar{p}$
collisions at $\sqrt{s}=1.96$~TeV over a wide range in $p_T^\gamma$,
$23<p_T^\gamma<300$ GeV.  This extends previous measurements in this
energy
regime~\cite{Abe:1994rr,Abbott:1999kd,Abazov:2001af,Acosta:2002ya,Acosta:2004bg}
to significantly higher values of $p_T^\gamma$.  Results from NLO pQCD
calculations agree with the measurement within uncertainties.

We thank W.~Vogelsang, J.P.~Guillet, E.~Pilon, and M.~Werlen for their
assistance with theoretical calculations.
%
We thank the staffs at Fermilab and collaborating institutions, 
and acknowledge support from the 
DOE and NSF (USA);
CEA and CNRS/IN2P3 (France);
FASI, Rosatom and RFBR (Russia);
CAPES, CNPq, FAPERJ, FAPESP and FUNDUNESP (Brazil);
DAE and DST (India);
Colciencias (Colombia);
CONACyT (Mexico);
KRF and KOSEF (Korea);
CONICET and UBACyT (Argentina);
FOM (The Netherlands);
PPARC (United Kingdom);
MSMT (Czech Republic);
CRC Program, CFI, NSERC and WestGrid Project (Canada);
BMBF and DFG (Germany);
SFI (Ireland);
Research Corporation,
Alexander von Humboldt Foundation,
and the Marie Curie Program.

\bibliography{paper}
\bibliographystyle{apsrev}
\end{document}